\documentclass{aa}
\usepackage{txfonts}
\usepackage{natbib}
\usepackage{graphicx}

\bibpunct[, ]{(}{)}{;}{a}{}{,}

\def\thestar{HD~3980}

\def\llm{{\sc LLmodels}}
\def\atl{{\sc ATLAS9}}

\def\logg{\log g}
\def\teff{T_{\rm eff}}

\def\kms{km/s}

\def\halpha{H$\alpha$}

\def\synth{{\sc Synth3}}
\def\synthmag{{\sc Synthmag}}

\def\elow{$E_{\rm low}$}
\def\loggf{$\log(gf)$}

\def\bl{$\langle B_{\rm l} \rangle$}

\def\vsini{$\upsilon\sin i$}

\def\b{|\mathrm{\mathbf{B}}|}

\def\Lsun{L_{\odot}}
\def\Rsun{R_{\odot}}

\begin{document}

\title{Multi-element Doppler imaging of the CP2 star HD~3980}

\author{N. Nesvacil\inst{1,7} \and T. L\"uftinger\inst{1} \and D. Shulyak\inst{2} \and M. Obbrugger\inst{1} \and 
        W. Weiss\inst{1} \and N.~A. Drake\inst{3,9} \and S. Hubrig\inst{4} \and T. Ryabchikova\inst{5} \and\\
        O. Kochukhov\inst{6} \and N. Piskunov\inst{6}, N. Polosukhina\inst{8}}
\offprints{D. Shulyak, \\
\email{denis.shulyak@gmail.com}}
\institute{
Universit\"at Wien, Institut f\"ur Astronomie, T\"urkenschanzstra{\ss}e 17, 1180 Wien, Austria \and
Institute of Astrophysics, Georg-August University, Friedrich-Hund-Platz 1, D-37077 G\"ottingen, Germany \and
Sobolev Astronomical Institute, St. Petersburg State University, Universitetski pr. 28, St. Petersburg 198504, Russia \and
Astrophysikalisches Institut Potsdam, An der Sternwarte 16, 14482 Potsdam, Germany \and
Institute of Astronomy, Russian Academy of Science, Pyatnitskaya 48, 119017 Moscow, Russia \and
Department of Physics and Astronomy, Uppsala University, Box 515, 751 20, Uppsala, Sweden \and
Department of Radiotherapy, Medical University of Vienna, W\"ahringer G\"urtel 18-20, 1090 Wien, Austria \and
Crimean Astrophysical Observatory, Nauchnyi, Crimea, Ukraine \and
Observat\'orio Nacional/MCT, Rua Gen. Jos\'e Cristino 77, 20921-400 Rio de Janeiro, Brazil
}

\date{Received / Accepted}

\abstract
{In atmospheres of magnetic main-sequence stars, the diffusion of chemical elements leads to a number
of observed anomalies, such as abundance spots across the stellar surface. }
{The aim of this study was to derive a detailed picture of the surface abundance distribution of the magnetic chemically peculiar star \thestar. }
{Based on high-resolution, phase-resolved spectroscopic observations of the magnetic A-type star
\thestar\, the inhomogeneous surface distribution of $13$
chemical elements (Li, O, Si, Ca, Cr, Mn, Fe, La, Ce, Pr, Nd,  Eu, and Gd) has been reconstructed. The INVERS12 code was used to invert the rotational variability in line profiles to elemental surface distributions.}
{Assuming a centered, dominantly dipolar magnetic field configuration, we find that Li, O, Mg, Pr, and Nd are mainly
concentrated in the area of the magnetic poles and depleted in the regions
around the magnetic equator. The high abundance spots of Si, La,
Ce, Eu, and Gd are located between the magnetic
poles and the magnetic equator. Except for La, which is clearly
depleted in the area of the magnetic poles, no obvious correlation with
the magnetic field has been found for these elements otherwise.
Ca, Cr, and Fe appear enhanced along the rotational equator and the area around the magnetic poles. 
The intersection between the magnetic and the rotational equator constitutes an exception, especially for Ca
and Cr, which are depleted in that region.}
{No obvious correlation between the theoretically predicted abundance patterns
and those determined in this study could be found. 
This can be attributed to a lack of up-to-date theoretical models, especially for
rare earth elements.}


\maketitle

\section{Introduction}
Chemically peculiar (CP2) stars are main-sequence objects
with abundance peculiarities such as strong over- or underabundance
of certain elements compared to the atmosphere of the Sun. Compared to their ``normal analogs'', large magnetic fields 
are frequently detected, with characteristic field strengths of the a few kG.

Because of their rather high effective temperatures (corresponding to spectral types B--F), relatively slow rotation, and the presence of 
strong magnetic fields, the mixing processes in atmospheres of Ap, stars are believed to be far less
efficient than in any other stars of the same spectral type. This opens a possibility for
slow dynamical processes to control the chemical structure of atmospheres, where particle diffusion
rules the spatial distribution of atoms of different species, depending on the balance between gravitational
and radiative forces \citep{michaud-1970}. As a result, elements become inhomogeneously distributed both vertically
and horizontally, which results in  vertical abundance stratification and surface abundance spots. These spots are observed through rotational 
line profile variations of respective elements.
Identifying links between abundance peculiarities, magnetic field configuration, and other atmospheric
parameters thus helps to put constraints on the efficiency of diffusion processes in low-density plasma.

\thestar\ ($\xi$~Phe or HR~183) is an A-type star belonging to the group of CP2 stars as classified by \citet{preston-1974}.
It was classified as an SrCrEu star by \citep{bidelman-macconnell-1973}. Among the magnetic stars of spectral type A (Ap stars), the so-called rapidly oscillating (roAp) stars  
show fast pulsations on typical time scales of 6 to 20 minutes \citep[see reviews by, e.g.,][]{kochukhov-2008,kurtz-2009}.
Although \thestar\ is observed close to the temperature range covered by roAp stars, 
no pulsations have been detected so far ~\citep{weiss-1979,martinez-kurtz-1994,elkin-2008}.
A magnetic field was first observed in \thestar\ by \citet{maitzen-1980}.
\thestar\ is a visual binary, with the secondary having an apparent
brightness in $V$ of $9.98$~mag \citep{weiss-1979}.

Because of its brightness and relatively fast rotation, \thestar\ is an ideal target for a detailed
study of surface abundance inhomogeneities. In this work we aim to reconstruct the abundance maps
based on phase-resolved, high-resolution, and high signal-to-noise spectra obtained with different instruments.
Dense phase and wavelength coverage allows abundance maps to be derived for $13$ chemical elements
(Li, O, Si, Ca, Cr, Mn, Fe, La, Ce, Pr, Nd,  Eu, and Gd), which presents one of the most complete
Doppler imaging studies to date.

\section{Observations}

\begin{table*}
	\caption{\textbf{Observations used for Doppler imaging.}}
\label{tab:obsreordered}
\begin{footnotesize}
\begin{center}
\begin{tabular}{cccccc}
\hline
\hline
HJD    & $\phi$      & Instrument & Resolution & SNR & Wavelength range [\AA]\\
\hline
2 452 173.975 & 0.064 & Mt Stromlo & 88 000 & 100 & 6180 - 7840\\
2 452 174.155 & 0.109 & Mt Stromlo & 88 000 & 100 & 6250 - 7840\\
2 452 170.991 & 0.309 & Mt Stromlo & 88 000 & 100 & 6700 - 6720\\
2 452 172.002 & 0.565 & Mt Stromlo & 88 000 & 100 & 6700 - 6720\\
2 452 172.201 & 0.615 & Mt Stromlo & 88 000 & 100 & 6700 - 6720\\
2 452 175.194 & 0.372 & Mt Stromlo & 88 000 & 100 & 6700 - 6720\\
2 452 178.180 & 0.128 & Mt Stromlo & 88 000 & 100 & 6700 - 6720\\
2 452 179.080 & 0.356 & Mt Stromlo & 88 000 & 100 & 6700 - 6720\\
2 452 180.087 & 0.611 & Mt Stromlo & 88 000 & 100 & 6700 - 6720\\
2 452 180.953 & 0.830 & Mt Stromlo & 88 000 & 100 & 6700 - 6720\\
2 452 181.071 & 0.860 & Mt Stromlo & 88 000 & 100 & 6700 - 6720\\
2 452 182.089 & 0.117 & Mt Stromlo & 88 000 & 100 & 6700 - 6720\\
2 452 182.192 & 0.143 & Mt Stromlo & 88 000 & 100 & 6700 - 6720\\
2 452 182.951 & 0.335 & Mt Stromlo & 88 000 & 100 & 6700 - 6720\\

2 453 298.518 & 0.643 & UVES 04 & 115 000 / 95 000 & 350 & 4800 - 6750\\
2 453 329.561 & 0.498 & UVES 04 & 115 000 / 95 000 & 250 & 4800 - 6750\\
2 453 329.558 & 0.499 & UVES 04 & 115 000 / 95 000 & 350 & 4800 - 6750\\
2 453 333.574 & 0.514 & UVES 04 & 115 000 / 95 000 & 300 & 4800 - 6750\\
2 453 333.577 & 0.515 & UVES 04 & 115 000 / 95 000 & 300 & 4800 - 6750\\

2 453 334.521 & 0.754 & HARPS & 115 000 & 150 & 3780 - 6900\\
2 453 334.592 & 0.772 & HARPS & 115 000 & 200 & 3780 - 6900\\
2 453 581.770 & 0.324 & HARPS & 115 000 & 200 & 3780 - 6900\\
2 453 582.798 & 0.584 & HARPS & 115 000 & 100 & 3780 - 6900\\
2 453 583.904 & 0.864 & HARPS & 115 000 & 200 & 3780 - 6900\\

2 453 631.707 & 0.961 & UVES 05 & 115 000 / 95 000 & 300 & 5735 - 9260\\
2 453 632.655 & 0.201 & UVES 05 & 115 000 / 95 000 & 300 & 5735 - 9260\\
2 453 632.659 & 0.202 & UVES 05 & 115 000 / 95 000 & 300 & 5735 - 9260\\
2 453 636.662 & 0.215 & UVES 05 & 115 000 / 95 000 & 350 & 5735 - 9260\\
2 453 636.665 & 0.215 & UVES 05 & 115 000 / 95 000 & 350 & 5735 - 9260\\
2 453 658.626 & 0.773 & UVES 05 & 115 000 / 95 000 & 250 & 5735 - 9260\\
2 453 658.629 & 0.774 & UVES 05 & 115 000 / 95 000 & 300 & 5735 - 9260\\
2 453 662.536 & 0.762 & UVES 05 & 115 000 / 95 000 & 300 & 5735 - 9260\\
2 453 663.552 & 0.020 & UVES 05 & 115 000 / 95 000 & 300 & 5735 - 9260\\
2 453 663.556 & 0.020 & UVES 05 & 115 000 / 95 000 & 300 & 5735 - 9260\\
2 453 666.747 & 0.828 & UVES 05 & 115 000 / 95 000 & 300 & 5735 - 9260\\
2 453 667.519 & 0.023 & UVES 05 & 115 000 / 95 000 & 300 & 5735 - 9260\\
2 453 669.637 & 0.559 & UVES 05 & 115 000 / 95 000 & 300 & 5735 - 9260\\

2 453 711.620 & 0.184 & HARPS & 115 000 & 200 & 3780 - 6900\\
2 453 712.606 & 0.433 & HARPS & 115 000 & 150 & 3780 - 6900\\
2 453 713.565 & 0.676 & HARPS & 115 000 & 150 & 3780 - 6900\\
2 453 714.622 & 0.943 & HARPS & 115 000 & 200 & 3780 - 6900\\
2 453 715.558 & 0.180 & HARPS & 115 000 & 200 & 3780 - 6900\\
2 453 716.587 & 0.441 & HARPS & 115 000 & 200 & 3780 - 6900\\

2 453 939.896 & 0.952 & UVES 06 & 115 000 / 95 000 & 400 & 4960 - 6990\\
2 453 941.939 & 0.468 & UVES 06 & 115 000 / 95 000 & 400 & 4960 - 6990\\



2 454 763.629 & 0.407 & FEROS 08 & 48 000 & 300 & 3560 - 9200\\

\hline
\end{tabular}
\end{center}
\end{footnotesize}
\end{table*}

Several of high-resolution
spectra were obtained for studying the chemical structures in \thestar\ . A FEROS spectrum was obtained with the $2.2$~m ESO telescope
at La Silla (Chile) under the agreement with the Observat\'orio
Nacional (Brazil). 
The Mt. Stromlo $74.3$-inch telescope and the coud\'e echelle spectrograph were used to obtain a series of spectra in September-October 2001,
as presented in \citep{2003IAUS..210P.D27P}.
Additionally, spectra acquired with UVES in 2004 (074.D-0392(A) Nesvacil et al.) and 2005 (076.D-0535(A)
Nesvacil et al.), HARPS data (074.C-0102(A) Hatzes et al., 075.C-0234(B)
Hatzes et al., 076.C-0073(A) Hatzes et al.), and UVES spectra from 2006
(077.D-0150(A) Kurtz et al.) were included in this analysis. The HARPS and UVES data from 2006 were
acquired via data mining and downloaded from 
the ESO archive\footnote{http://archive.eso.org/eso/eso\_archive\_adp.html}.

A list of all observations is displayed in Table~\ref{tab:obsreordered}. Since most of
the available spectra obtained with Mt~Stromlo are dedicated to the \ion{Li}{i} feature at $6708$~\AA,
their spectral range is restricted to this wavelength region.

Rotational phases were computed using ephemeris from \citet{maitzen-1980}, where zero phase corresponds to the prime minimum in the
$v$-filter curve of the Str\"omgren system (positive extrema of the longitudinal magnetic field):

\begin{eqnarray}
 JD(\mathrm{Prime \: min. \: in \: v-filter}) & =  & 2442314.48\nonumber \\
                                     &\pm & 0.04 + 3\fd9516 \pm 0.0003 .
\end{eqnarray}

\begin{figure}
\centerline{\includegraphics[width=0.8\hsize]{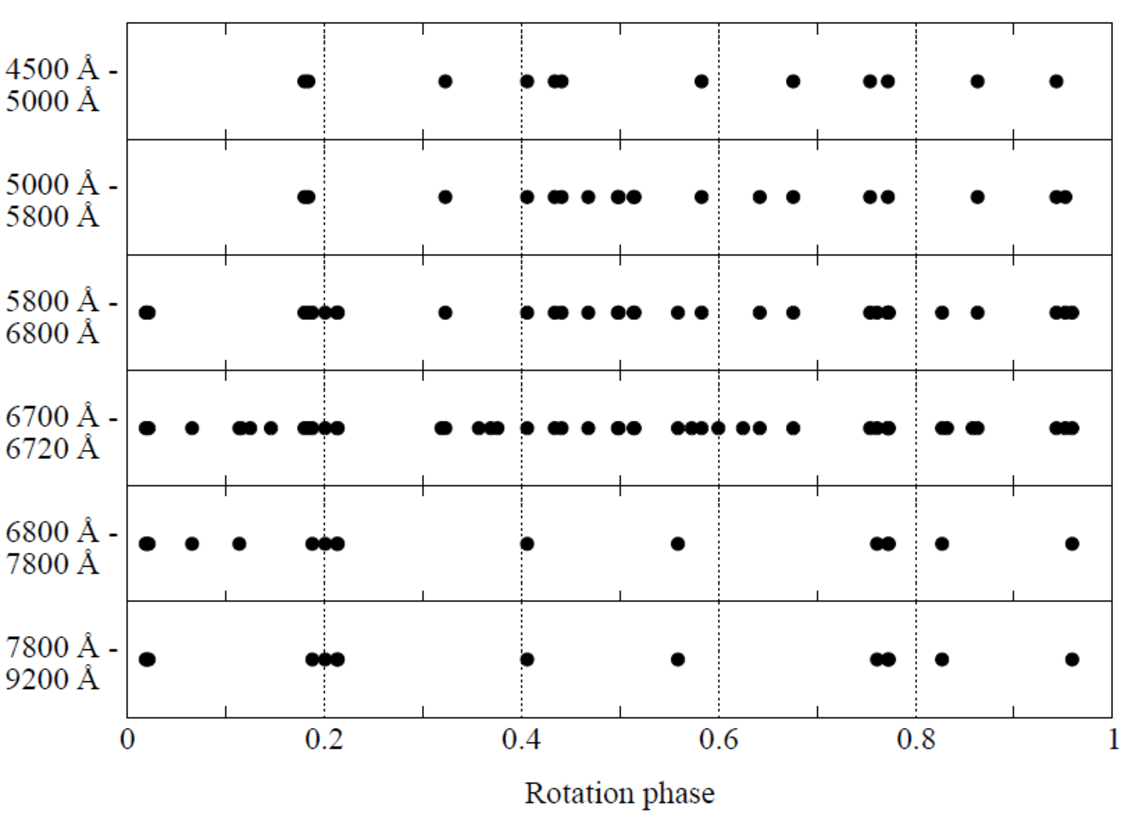}}
\caption{Phase coverage of the data used for Doppler imaging. The different
panels show the phase coverage of the several wavelength regions, which changes
due to the combination of spectra from various spectrographs and settings (see text).}
\label{fig:phase}
\end{figure}

The relatively large rotational period error, estimated by \citet{maitzen-1980} from the visual examination of light curves,
suggests significant uncertainties in the phase calculation for our spectroscopic data, making it problematic to relate DI maps 
to the magnetic and photometric variations. However, \citet{elkin-2008} shows that the phasing of the new and old magnetic 
field measurements agrees to within 0.05 of the rotational period. This implies a period error of $\approx6\times10^{-5}$, which 
is small enough to be negligible for our analysis and interpretation of the mapping results. Our own
estimates of the period accuracy led to a similar value of $\approx4\times10^{-5}$.

The resolution of the spectra ranges from $48\,000$ to $115\,000$, as shown in
Table~\ref{tab:obsreordered}. For UVES spectra the resolution varies from $95\,000$ for the blue to $115\,000$ for 
the red arms, when using the narrowest slit width available. Depending on the setting, which is different for
all the UVES runs, gaps in the spectral coverage occur at different wavelengths. 
by combining all available observations the resulting 
phase coverage has a maximum phase gap of $\Delta\phi=0.2$, which is visualized in Fig.~\ref{fig:phase}.

Basic data reduction was performed with pipelines of the corresponding
instruments\footnote{http://archive.eso.org/eso/eso\_archive\_adp.html}$^,$\footnote{http://www.eso.info/sci/data-processing/software/gasgano/}.
For the Mt Stromlo and FEROS data, the reduction was done
with the Image Reduction and Analysis Facility (IRAF, \citet{tody}) and standard recipes.

Additional Str\"omgren and Johnson photometry was used and is summarized in Table ~\ref{tab:param}.

\section{Basic methods}
\label{sec:methods}

\begin{figure}
\centerline{\includegraphics[width=\hsize]{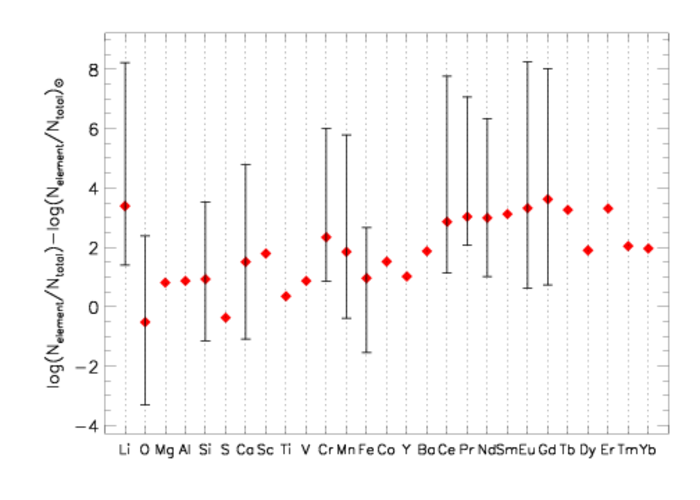}}
\caption{Elemental abundances of \thestar\ relative to the Sun. Vertical bars are attached to mapped elements and
correspond to maximum and minimum surface abundances as shown in abundance maps (see Sect.~\ref{sec:results} for more details,
Figs.~\ref{fig:maps1},~\ref{fig:maps2}, and~\ref{fig:maps3}). 
Reference solar abundances are from \citet{asplund-2005}.}
\label{fig:abn}
\end{figure}

Over the course of this work two different model atmosphere codes were
used iteratively. As a starting point, \atl\ model atmospheres \citep{a9-1,a9-2}were computed because the use of 
opacity tables allows for very fast calculations. An initial abundance
analysis was performed based on equivalent widths and individual line profile fitting
in the case of blended lines. For further analysis \llm\ model atmospheres \citep{llm}
were considered a more appropriate choice for a star with a rather complex abundance
pattern such as for \thestar. The \llm\ code was used to construct a final model atmosphere
with the individual abundances presented in Fig.~\ref{fig:abn}.
If elements were found to be inhomogeneously distributed across the stellar surface, average homogeneous abundance values were 
implemented in the model atmosphere calculations.

To calculate synthetic spectra the \synth\ code described by \citet{synthmag2007} was used.
The list of atomic lines used in our study was taken from the VALD data base \citep{vald1,vald2}.
Further information on the final linelist used for reconstructing
abundance maps is given in the Table~\ref{tab:linelist}.

\begin{table}
\caption{Table of elements and spectral lines from VALD used for mapping \thestar.}
\label{tab:linelist}
\begin{center}
\begin{tabular}{lcccccc}
\hline
\hline
Species    & $\lambda$, \AA  & \loggf & \elow, eV &  Blended with\\
           &                  &        &             & \\
\hline
\ion{Mn}{ii} & 4737.9440 & -2.940 & 6.128 &\\
\ion{Mn}{ii} & 4738.2900 & -1.876 & 5.380 &\\
\hline
\ion{Ce}{ii} & 5518.3580 & -2.030 & 0.553 &\\
\ion{Ce}{ii} & 5518.4570 & -2.000 & 1.482 &\\
\ion{Ce}{ii} & 5518.4890 & -0.670 & 1.155 &\\
\ion{Ce}{ii} & 5518.6030 & -2.430 & 1.079 &\\
\hline
\ion{La}{ii} & 5880.6330 & -1.830 & 0.235 & Cr\\
\hline
\ion{Gd}{ii} & 5855.2150 & -1.020 & 1.598 &\\
\ion{Gd}{ii} & 6004.5590 & -0.780 & 1.659 &\\
\hline
\ion{Fe}{i}  & 6065.4820 & -1.530 & 2.608 &\\
\ion{Fe}{i}  & 6230.7230 & -1.281 & 2.559 & Ce\\
\hline
\ion{Cr}{ii} & 6176.9810 & -2.887 & 4.750 &\\
\ion{Cr}{ii} & 6177.2060 & -3.093 & 6.686 &\\
\ion{Cr}{ii} & 6379.7920 & -3.767 & 4.497 & Fe, Gd\\
\ion{Cr}{ii} & 6226.6380 & -3.035 & 4.756 & Ce\\
\hline
\ion{Eu}{ii} & 6437.6400 & -0.320 & 1.320 &\\
\hline
\ion{Ca}{i}  & 6439.0750 & 0.390  & 2.526 &\\
\hline
\ion{Li}{i} & 6707.7561 & -0.440 & 0.000 & Ce, Pr, Sm\\
\ion{Li}{i} & 6707.7682 & -0.239 & 0.000 &\\
\ion{Li}{i} & 6707.9066 & -0.965 & 0.000 &\\
\ion{Li}{i} & 6707.9080 & -1.194 & 0.000 &\\
\ion{Li}{i} & 6707.9187 & -0.745 & 0.000 &\\
\ion{Li}{i} & 6707.9200 & -0.965 & 0.000 &\\
\hline
\ion{Nd}{iii} & 6145.0677 & -1.330 & 0.296 & Si, Ca\\
\hline
\ion{O}{i} & 7771.9413 & 0.369 & 9.146 & Cr\\
\ion{O}{i} & 7774.1607 & 0.223 & 9.146 & Nd, Gd\\
\ion{O}{i} & 7775.3904 & 0.001 & 9.146 &\\
\hline
\ion{Pr}{iii} & 7781.9830 & -1.210 & 0.000 & Fe, Gd\\
\hline
\ion{Si}{i} & 7932.3480 & -0.468 & 5.964 &\\
\ion{Si}{i} & 7944.0010 & -0.293 & 5.984 & Cr\\
\hline
\end{tabular}
\end{center}
\end{table}

The observed line profiles were inverted with the \textsc{INVERS12} code \citep{kochukhov-2004}.
The code uses a regularized minimization algorithm and can treat many individual lines and
their blends simultaneously. A detailed description of the software and some practical applications are 
given by, e.g., \citet{piskunov-rice-1993},\citet{lueftinger-2010}, and \citet{kochukhov-2004}.

\section{Results}
\label{sec:results}

\subsection{Fundamental parameters}

The TempLogG$^{\rm TNG}$ program \citep{kaiser-2006} was used to derive fundamental parameters from photometric calibrations 
\citep{moon-dworetsky-1985}. The available Johnson $V$ and Str\"omgrem photometry yields $\teff=8340$\,K and $\logg=4.3$, but
without any error estimation. 


This initial estimation was followed by an attempt to determine $\teff$ and $\logg$ by fitting the 
\halpha\ line with a synthetic line profile, which was, unfortunately, not successful owing to large differences between the
\halpha\ line profiles of the various spectra. No trend that correlates those changes to
the rotational period was found. It is possible that these variations are caused by the reduction pipelines, the 
continuum fitting procedure, or the curvature of the
echelle spectra.

As reported in \citet{obbrugger-2008}, the temperature was previously
estimated to be $8500$\,K. This value was based on equivalent width measurements
of several \ion{Fe}{i} and \ion{Fe}{ii} lines, which also yielded $\logg=4.0$. 
Previous publications give several rather
similar values for the temperature, e.g., $8000\pm250$\,K \citep{drake-2004},
$8270\pm200$\,K \citep{k-and-b-2006}, or $8240\pm300$\,K \citep{hubrig-2007}. 
The $\logg$ values published so far are $4.04\pm0.05$ from \citet{drake-2004},
$4.05\pm0.09$ in \citet{hubrig-2007}, and $4.0\pm0.2$ by \citet{elkin-2008}.
Consequently, $\teff=8300\pm250$\,K and $\logg=4.0\pm0.2$ were adopted for the Doppler 
mapping procedure in this work.

The absolute magnitude of the star was calculated using the new reduction of Hipparcos data. Stellar luminosity and radius were computed
with the bolometric correction from \citet{landstreet-2007}.



The value of the projected rotational velocity \vsini\ was determined by
comparing of the synthetic spectra with observations of magnetically insensitive
iron lines. This method led to a \vsini$=22.5\pm2$~\kms.
Combination of the measured \vsini, period, and radius results in
an inclination angle of $i < 44$ degrees. 
\begin{figure}
\centerline{\includegraphics[width=0.8\hsize]{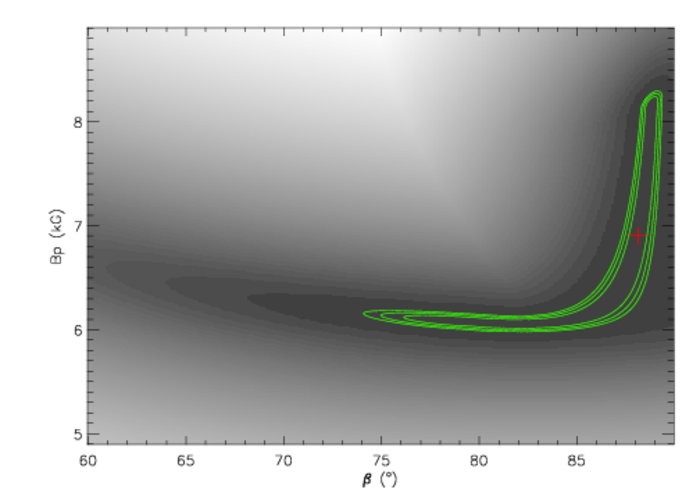}}
\caption{Chi-square surface as a function of dipolar strength ($B_{\rm p}$) and magnetic obliquity ($\beta$). The cross marks the best-fit
solution, while the contours 
	correspond to the $1$, $2$, $3$ sigma limits. The uncertainty of the inclination angle is taken into account implicitly. 
For each $B_{\rm p}$-$\beta$ pair the lowest chi-square was recorded, varying inclination within its $\pm 1$ sigma range in ${\rm sin}~i$.}
\label{fig:mag}
\end{figure}
Magnetic field measurements for this star have been reported by \citet{maitzen-1980}, 
\citet{hubrig-2006} and \citet{elkin-2008}.
Since only measurements of the longitudinal magnetic field \bl\ are presently available, 
the magnetic field geometry is assumed to be a centered dipole (which is usually
a good approximation in the case of magnetic A stars). 
Using a relation between the maximum and minimum of the longitudinal magnetic field 
and inclination angle $i$ \citep[see][]{preston-1967}, an angle $\beta=88^\circ$ was therefore found between the line-of-sight and the 
rotation axis of the star.
From the fit to the measurements of the mean longitudinal field \citet{elkin-2008} obtained an amplitude of $1904$~G. 
The positive magnetic pole is located near the zero rotational phase  (light minimum). 

Because of the relatively small error
in the period estimated by \citet{elkin-2008}, we do not expect large changes in the position of the magnetic extrema. This conclusion
also follows from Fig.~2 of \citet{elkin-2008}, which illustrates a good match between the two magnetic measurements from
different epochs (1980 and 2006) folded with the same period. Last but not least, the most recent magnetic measurements cited in \citet{elkin-2008} 
and our spectroscopic measurements overlap in time, thus justifying the use of the ephemeris from \citet{maitzen-1980}.
Finally, using the relation from \citet{preston-1967}, which combines the amplitude of the longitudinal magnetic field
variation, inclination angle, magnetic angles, and limb-darkening coefficient
(assuming $u=0.5$), we
obtain the polar strength of the dipolar field $B_{\rm p}\approx7$~kG. Taking the uncertainty of the 
inclination angle into account, the 1 sigma confidence ranges of these parameters are $B_{\rm p}=6.01-8.25$~kG and $\beta=76^\circ-89^\circ$. Figure~\ref{fig:mag}
illustrates the analysis of the $B_{\rm z}$ measurements for \thestar, showing
the chi-square surface as a 
function of dipolar strength ($B_{\rm p}$) and magnetic obliquity ($\beta$).

\begin{table}
\caption{Fundamental and atmospheric parameters of \thestar\ used for Doppler imaging.}
\label{tab:param}
\begin{center}
\begin{tabular}{lcc}
\hline
\hline
Parameter & Value& Reference\\
\hline
$H_\beta$& 2.870&(1)\\
$b-y$&$0.070 \pm 0.007$&(1)\\
$m1$&$0.296 \pm 0.007$&(1)\\
$c1$&$0.748 \pm 0.022$&(1)\\
$V$&$5.71 \pm 0.005$&(1)\\
\hline
$\pi$, mas& $14.91 \pm 0.35$&(2)\\
$d$, pc&$67.07 \pm 1.57$&(2)\\
$BC$, mag & $0.6 \pm 0.07$&(3)\\
$M_{\rm bol}$, mag & $1.64 \pm 0.079$&\\
$\log(L/\Lsun)$ & $1.24 \pm 0.04$& \\
$R/\Rsun$ & $2.04 \pm 0.26$&\\
\hline
\vsini, \kms & $22.5 \pm 5$&\\
$i$, $^\circ$ & 60 (14-90)&\\
$\beta$, deg & 88.1 (76-89)&\\
$B_{\rm p}$, kG & 6.9 (6.0-8.3)&\\
\hline
$\teff$, K & $8300 \pm 250$&\\
$\logg$, cgs & $4.0 \pm 0.2$&\\
\hline
\multicolumn{3}{l}{\footnotesize{Photometric measurements were taken from 1:}}\\ 
\multicolumn{3}{l}{\footnotesize{\citet{hauck-mermilliod-1998}, 2: \citet{van-Leeuwen-2007},}}\\ 
\multicolumn{3}{l}{\footnotesize{3: \citet{landstreet-2007}.}}\\
\end{tabular}
\end{center}
\end{table}

All photometric measurements and resulting fundamental parameters of \thestar\ are summarized in Table~\ref{tab:param}.

\subsection{Surface abundance distributions}

\begin{figure*}
\centerline{\includegraphics[width=0.8\hsize]{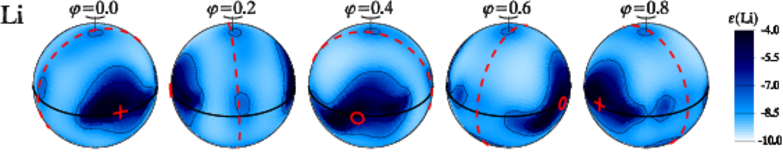}}
\centerline{\includegraphics[width=0.8\hsize]{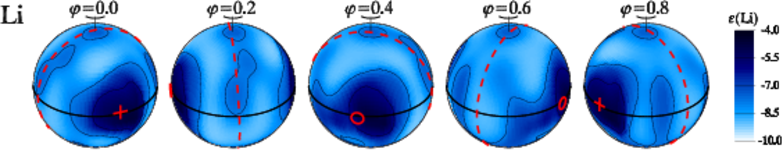}}
\centerline{\includegraphics[width=0.8\hsize]{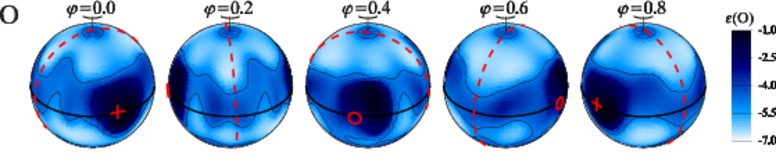}}
\centerline{\includegraphics[width=0.8\hsize]{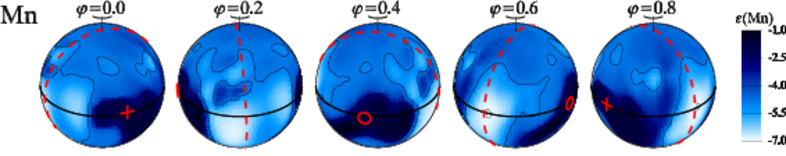}}
\centerline{\includegraphics[width=0.8\hsize]{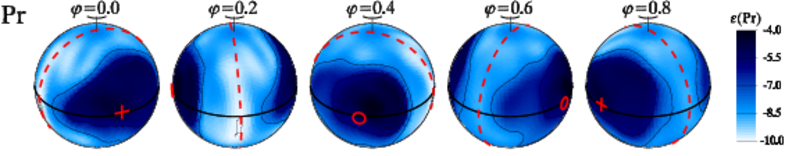}}
\centerline{\includegraphics[width=0.8\hsize]{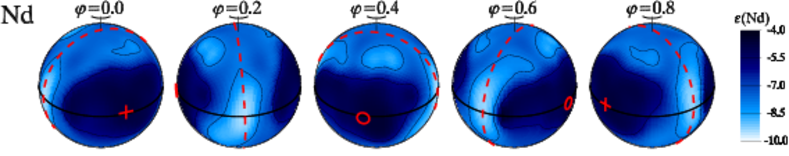}}
\caption{Abundance distributions of Li, O, Mn, Pr, and Nd at the surface of \thestar\ obtained from the
lines listed in Table~\ref{tab:linelist}. The star is shown at five rotational phases. Darker areas correspond
to the higher elemental abundances, the corresponding scale is given to the right of each panel 
and contours with equal abundances are plotted with steps of $1$~dex. 
The circle and the cross indicate the position of the negative and the positive magnetic poles, respectively.
This figure represents elements of \textit{Group~1} with enhanced spots centered on the magnetic poles.
Two Li maps correspond (from top to bottom) to the cases of simultaneous mapping 
of \ion{Li}{i} and \ion{Pr}{iii} blend at $6707$~\AA,
and additional implementation of \ion{Pr}{iii}~$7781.983$~\AA\ line (see text).
Abundances are given in $\log(N_{\rm el}/N_{\rm total})$.}
\label{fig:maps1}
\end{figure*}

\begin{figure*}
\centerline{\includegraphics[width=0.8\hsize]{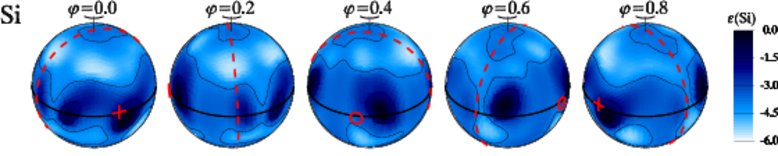}}
\centerline{\includegraphics[width=0.8\hsize]{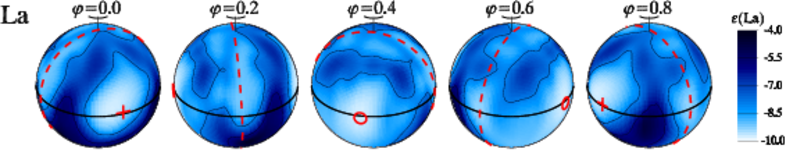}}
\centerline{\includegraphics[width=0.8\hsize]{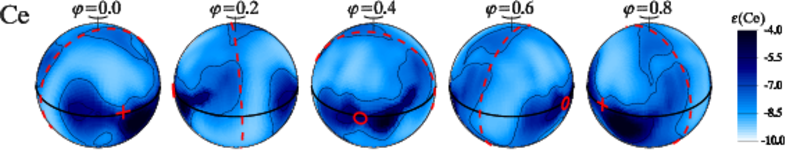}}
\centerline{\includegraphics[width=0.8\hsize]{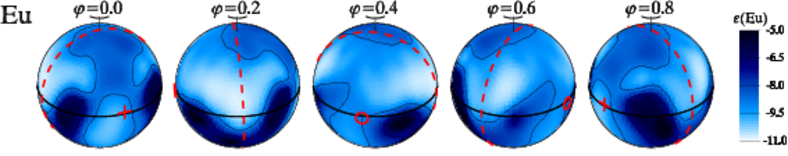}}
\centerline{\includegraphics[width=0.8\hsize]{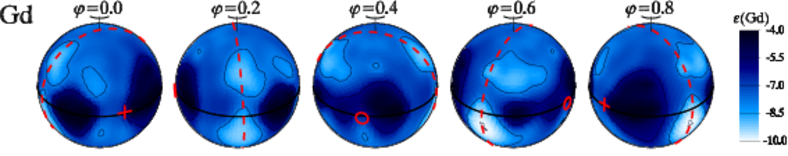}}
\caption{Same as on Fig.~\ref{fig:maps1}, but for Si, La, Ce, Eu, and Gd.
This figure represents elements of \textit{Group~2} with spots of higher abundance between the magnetic poles and
the magnetic equator.}
\label{fig:maps2}
\end{figure*}

\begin{figure*}
\centerline{\includegraphics[width=0.8\hsize]{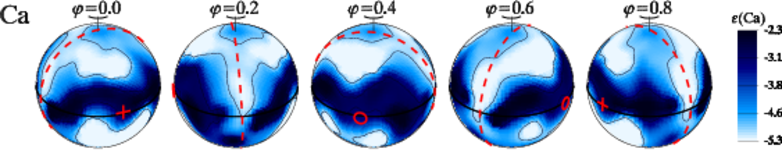}}
\centerline{\includegraphics[width=0.8\hsize]{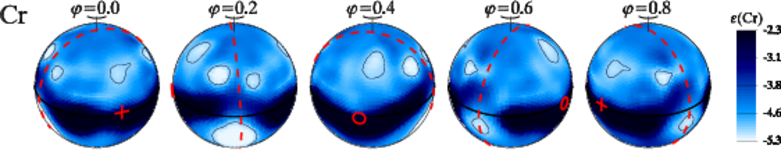}}
\centerline{\includegraphics[width=0.8\hsize]{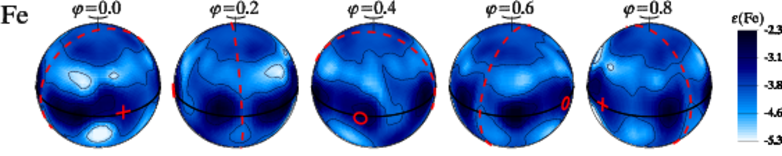}}
\caption{Same as on Fig.~\ref{fig:maps1}, but for Ca, Cr, and Fe.
This figure represents elements of \textit{Group~3} with overabundances in the area of the magnetic poles along
the rotational equator.}
\label{fig:maps3}
\end{figure*}

\begin{figure*}
\begin{center}
\includegraphics[height=0.9\vsize,width=0.17\hsize]{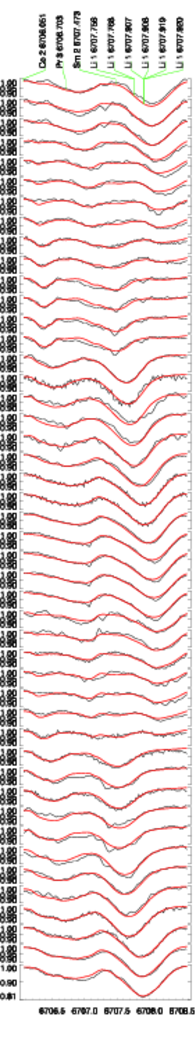}
\includegraphics[height=0.9\vsize,width=0.17\hsize]{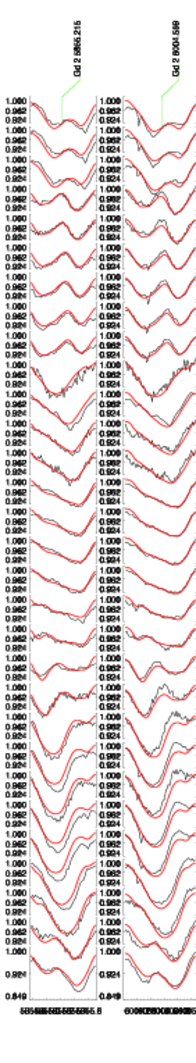}
\includegraphics[height=0.9\vsize,width=0.17\hsize]{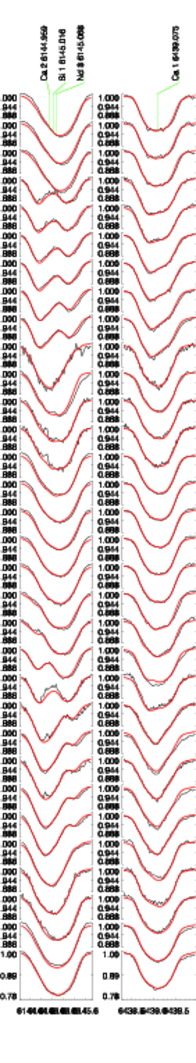}
\includegraphics[height=0.9\vsize,width=0.17\hsize]{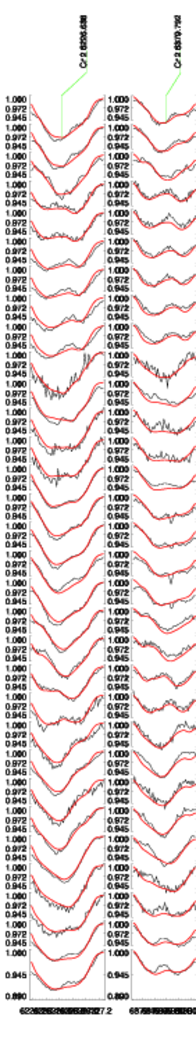}
\includegraphics[height=0.9\vsize,width=0.17\hsize]{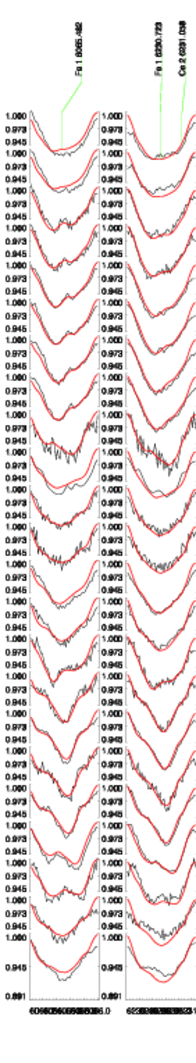}
\end{center}
\caption{Examples of observed and predicted phase-dependent profiles of selected elements computed from the respective 
abundance maps shown in Figs.~\ref{fig:maps1},~\ref{fig:maps2}, and~\ref{fig:maps3}.}
\label{fig:profs}
\end{figure*}

In the following we give a short overview of the results for each individual element that has been mapped.
Abundance maps are combined in three groups according to the distribution pattern and its link
to the assumed magnetic field geometry. These maps are presented in Figs.~\ref{fig:maps1},~\ref{fig:maps2}, 
and \ref{fig:maps3} (see Sect.~\ref{sec:discussion}). Examples of the theoretical fits to phase-resolved profiles of
some selected lines are presented in Fig.~\ref{fig:profs} for illustration.
When comparing the magnitude of over- or underabundance of some elements we assume reference solar abundances 
from \citet{asplund-2005}. In all other cases (if not explicitly stated), when referring to the abundances 
of particular surface region, enhancements or depletion is discussed relative to the mean
surface abundance.

\subsubsection*{Lithium}
This map has the highest phase coverage ever
used for Doppler imaging of an A-type star.
Forty-six phases were available to reconstruct the
abundance variations of Li (see Fig.~\ref{fig:phase}).
\ion{Pr}{iii} at $6706.703$~\AA\ was varied simultaneously,
because the overlap in wavelength occurred when Li and Pr spots
were visible near the stellar limb. The
enhancements of Li at the magnetic poles
have an overabundance of $\approx5$~dex compared to the Sun. The solar abundance of Li
	is $\approx2$~dex lower than the initial cosmic abundance, \citep{baumann-2010}.

An additional parallel implementation of \ion{Pr}{iii} at $7781.983$~\AA\ leads
to a more extended concentric Li spot around the
magnetic poles with the same overabundance of
$\approx5$~dex (see Fig.~\ref{fig:maps1}, second plot from top). The enhancement at the negative pole was
slightly shifted towards a higher longitude, but resulting in slightly worse
fit to the \ion{Pr}{iii} line at $6706.703$~\AA. This could be due
to influences of the \ion{Sm}{ii} line at $6707.473$~\AA\ or
\ion{Ce}{ii} $6706.051$~\AA\ and $6708.099$~\AA\ lines that were only accounted for
as having a constant abundance over the whole stellar surface \citep[see, e.g.,][]{2005IAUS..228...89D}.

\subsubsection*{Oxygen}
The surface abundance pattern of oxygen was reconstructed
using the triplet at $7770$~\AA. The line
profile variations are very pronounced and reproduced well
in the area of these three O lines. Blends of Nd and Gd were included with  
 constant abundance contribution. Like Li, O is concentrated
in two spots at the magnetic poles, whereas the
enhancement at the negative pole is also shifted
towards a higher longitude. Around the magnetic
field equator oxygen appears rather depleted.
Oxygen abundance maps have so far been reported for the hotter Ap star $\epsilon$~UMa \citep{1997A&A...326..988R} and for 
HD~83368 (HR~3831), with a comparable temperature 
to \thestar\ \citep{kochukhov-2004}. It is interesting to note that the distribution of
oxygen abundance in \thestar\ is completely different from what
is observed in the two other stars, where oxygen is strongly
concentrated in the region of the magnetic equator. 

\subsubsection*{Silicon}
In order to derive the abundance
pattern of Si, two \ion{Si}{ii} lines at $7932.348$~\AA\ and $7944.001$~\AA\ were combined, where the weak
\ion{Cr}{ii} $7944.527$~\AA\ blend was implemented by assuming a
constant abundance over the rotation period.
Our results indicate that there are several spots of overabundance
around the rotational equator in the area between
the magnetic poles and the magnetic equator.

\subsubsection*{Calcium}
The Ca surface distribution was obtained by combining
the unblended \ion{Ca}{i} line at $6439.075$~\AA\ with the 
\ion{Nd}{iii}, \ion{Ca}{ii}, and \ion{Si}{i} blend at $6145$~\AA.
Since all elements contribute strongly to the
blend, their distribution was determined simultaneously
in this calculation. The resulting surface
image of Ca shows an overabundance in
the region of the rotational equator and in the
area of the magnetic poles of more than $2$~dex
compared to the Sun and displays approximately
solar abundance around the magnetic equator.

\subsubsection*{Chromium}
Chromium shows a very distinct structure of an
overabundance around the rotational equator
broken by an underabundance around the magnetic
equator. To minimize the effects of possible vertical abundance stratification 
(frequently found in roAp stars, \citet{hd24712}), two lines of the same ion
with the similar excitation energy were used (\ion{Cr}{ii} $6226.638$~\AA\ and \ion{Cr}{ii} $6379.792$~\AA).

\subsubsection*{Manganese}
Two lines of \ion{Mn}{ii} at $4737.944$~\AA\ and $4738.290$~\AA\ were used.
Due to the spectral coverage in this region, only
12 phases are available, but the phases are evenly spaced over the rotational cycle. The
surface distribution of manganese is characterized
by two patches of enhanced abundance
around the magnetic poles and depletion at
the intersections of the magnetic and rotational
equators.

\subsubsection*{Iron}
Two \ion{Fe}{i} lines at $6065.482$~\AA\ and $6230.723$~\AA\ have very
similar excitation energy and are practically
unblended. The \ion{Ce}{ii} line at $6231.058$~\AA\ was
taken into account by assuming a constant abundance.
Iron is distributed in high-contrast spots
of overabundance around the rotational equator,
and seems to accumulate at the magnetic
poles rather than at the magnetic equator.

\subsubsection*{Lanthanum}
The most remarkable features of the surface
distribution of La are the depletion
regions around the magnetic poles. The regions
with higher abundance are located in the area
between the magnetic poles and the magnetic
equator, and slightly shifted towards the southern
hemisphere.

\subsubsection*{Cerium}
The map resulting from the inversion of the \ion{Ce}{i}
lines at $5518$~\AA\ shows enhanced abundances
in the region between the magnetic poles and
the magnetic equator. The area around the
magnetic equator itself is depleted.

\subsubsection*{Praseodymium}
The result of the inversion of \ion{Pr}{iii} $7781.983$~\AA\ shows that two spots are observed
around the magnetic poles with an 
abundance increase of more than $5$~dex compared to the solar
value. Also the area of lowest
abundance around the magnetic equator is more
than $1$~dex higher than for the Sun.

We found that the combination of the Li line at $6708$~\AA\ with the \ion{Pr}{iii} $7781.983$~\AA\ line results 
in a very similar distribution that shows two patches in
the area of the magnetic poles, but their centers are
shifted by $+20^\circ$ in longitude. The spot size, as well
as the abundance value, agree, which
is also valid for the depletion surrounding the
magnetic equator. As mentioned before, the
praseodymium line contributing to the Li-Pr
blend at $6708$~\AA\ is rather weak. 
We therefore kept only the map resulting from the 
\ion{Pr}{iii} $7781.983$~\AA\ line.

\subsubsection*{Neodymium}
Neodymium was~--~similar to Ca~--~derived
from the combination of the \ion{Ca}{ii}, \ion{Si}{i},
and \ion{Nd}{iii} blend at $6145$~\AA\ and \ion{Ca}{i} line
at $6439.075$~\AA; however, the contribution of \ion{Ca}{ii} and \ion{Si}{i}
appeared to be weak. The map shows a depletion
surrounding the magnetic equator. Similar to Pr, two concentric areas of higher
abundance are centered on the magnetic poles.

\subsubsection*{Europium}
Just like Ce, the enhanced regions of Eu
are situated between the magnetic
equator and the magnetic poles. The two main
spots are shifted towards higher longitudes relative
to the magnetic poles.

\subsubsection*{Gadolinium}
Gadolinium shows enhanced spots in the area
of the rotational equator. Around the magnetic
equator, several spots of lower abundance are
visible. The enhanced areas are located between
the magnetic poles and the magnetic equator.
This surface abundance distribution was derived
by combining the two \ion{Gd}{ii} lines at $5855.215$~\AA\ and $6004.559$~\AA.

\section{Discussion}
\label{sec:discussion}

\subsection{Magnetic spectrum synthesis and abundance maps}

In the present Doppler imaging analysis we used a nonmagnetic spectrum synthesis
for computating the line profiles of respective elements. However,
with a polar magnetic field strength of up to $7$~kG and magnetic angle $\beta=88^\circ$, the mean
surface magnetic field modulus ranges from $4.5$~kG to $5.4$~kG during the rotation period.
Such a magnetic field is strong enough to influence the line profiles , so that 
its possible impact on the abundance maps has to be discussed.
 
It is known that the formation of the \ion{Li}{i} $6707.473$~\AA\ line
is subject to the Paschen-Back effect \citep[see][]{kochukhov-pb-2008,stift-pb-2008},
which becomes noticeable around $3$~kG. However, we did not account for this in the present study
because a) modeling the Paschen-Back effect is a non-trivial task that requires special 
software and complex techniques; b) the true geometry of the global magnetic field of \thestar\ is not known, i.e. its deviation
from the simple dipole would introduce additional uncertainties in the line profile fitting with magnetic
spectrum synthesis, and c) the star's rather high \vsini$=22.5$~\kms\ smears out the line profile shape considerably.

To estimate the impact of neglecting the
magnetic field in spectrum synthesis on the
derived abundances, especially of strongly overabundant elements, we used the \synth\ and \synthmag\ codes \citep{synthmag2007} to
calculate two sets of synthetic spectra for the
lines of Fe, Cr, Eu, and Gd which are most sensitive to the magnetic field
i) without magnetic field, and ii) considering a magnetic field modulus of $5$~kG. 
For both cases synthetic profiles were fit
to observed lines in a rotation phase close to maximum abundance. We found that including the
Zeeman splitting in the presence of a magnetic field leads to an abundance decrease
of $0.10-0.15$~dex for Fe and Cr. For the Eu line, in addition to the
Zeeman splitting, isotopic splitting and hyperfine structure 
were also taken into account in magnetic spectrum synthesis. 
A terrestrial Eu isotopic composition was adopted. 
Due to this complex pattern of the  \ion{Eu}{II} $\lambda$~$6437$ line used for mapping, the
abundance decreased by $0.65$~dex compared to the simplified, nonmagnetic case.
The highest abundance difference was derived for Gd. A large Land\'e factor 
and a complex Zeeman pattern meant its abundance in the magnetic case had to be decreased by $1.1$~dex to fit observations.
Except for Eu and Gd, the upper abundance limits of the maps computed with
INVERS12 are not very affected by neglecting the magnetic field. Lower
abundance limits are not influenced as much for any of the lines used for mapping.

Finally, it is important to understand that more accurate work could be done by applying
a magnetic Doppler imaging when both distributions of the magnetic field and abundances are mapped simultaneously.
This, of course, will become possible once additional polarimetric observations are available.

\subsection{Surface patterns and assumed magnetic field geometry}

As noted in Sect.~\ref{sec:results}, the single-wave variations in the longitudinal magnetic field of \thestar\ 
can correspond to the magnetic dipole with the obliquity angle $\beta=88^\circ$. The true magnetic field
configuration can, of course, be more complex. But since
no phase-resolved observations of additional magnetic parameters (e.g. field modulus $\b$) 
are available, it is currently impossible to determine the magnetic field topology precisely.
Nevertheless, assuming that the magnetic field is dominated by a dipolar-like configuration 
(which is commonly found in other A-type magnetic stars), we can look for a link between the
global magnetic field geometry and elemental abundance distributions.

The elements displayed in Fig.~\ref{fig:maps1} show more or less circular regions of higher abundance around the
magnetic poles (Group~1). Especially around the negative magnetic pole, a shift in the
enhanced spot towards higher longitudes is visible. Besides Pr and Nd, 
lighter elements such as Li, O, and Mn~--~an iron peak element~--~are also distributed in the same way. 
The other apparent feature of this group is lower abundance around the magnetic equator.

The next set of elements (Group~2), which are displayed in Fig.~\ref{fig:maps2} shows high
abundance regions placed mainly between the magnetic poles and magnetic equator.
Lanthanum even avoids the magnetic poles and shows a concentric region
of underabundance around them (in contrast to, e.g., O or
Pr). The region surrounding the magnetic equator, on the other
hand, is dominated by La-depleted areas.

The final Group~3, displayed in Fig.~\ref{fig:maps3}, is not as clearly defined, so it 
cannot be assigned to any of the other groups. Ca, Cr, and Fe
are overabundant along the rotational equator and the area of the magnetic
poles, but not in circular spots as in the case of, e.g., Li or Nd. The region
of the magnetic equator is depleted for all those elements.

Such a distinct separation of the mapped elements into the three groups described above 
can only be explained once the interplay between rotation, diffusion, and magnetic field
is correctly modeled in recent and/or future diffusion models.

\subsection{Comparison with diffusion models}
Assuming that the magnetic field geometry is a centered dipole, simple structures 
like rings or caps are expected, according to \citet{michaud-1981}. Caps
around the magnetic poles are found for Li, O, Mn, Pr, and Nd with depletion 
around the magnetic equators. The other 8 investigated elements have
a more complicated structure. None of the analyzed elements show enhancements 
around the magnetic equator. But, e.g., Si should be overabundant in
areas of close to horizontal magnetic field lines, as stated by \citet{alecian-vauclair-1981},
\citet{michaud-1981}, or \citet{vauclair-1979}.
The enhanced spots of Si are situated between the magnetic equator and
the magnetic poles. This might suggest that the diffusion process is still
going on.

To explain surface abundance patterns of Ap
stars, a further development of the diffusion theory 
was proposed by \citet{babel-1992,babel-1993,babel-1995}. It is claimed that additional modifications
of the outer boundary condition by a small and inhomogeneous mass outflow could considerably modify the surface abundance structure.
Ambipolar diffusion of hydrogen in magnetized plasma can also affect surface abundance patterns compared to the predictions
of the initial parameter-free diffusion theory where no mass loss or ambipolar diffusion were considered.
In particular, for stars with a $\teff$ of more than $10\,000$~K, a mass loss due to
radiative pressure is predicted for selected metals \citep{babel-1995}. Such weak
metal winds could produce surface inhomogeneities with varying contrast
between the elements.
Unfortunately, only a few of the elements investigated in this work were the 
subject of diffusion calculations, which are so far applicable to a limited
range of $\teff$ and $\logg$ values.

The Group~2 of elements (e.g. Fig.~\ref{fig:maps2}) especially shows asymmetric
distributions and variations of the abundance pattern from one element to
the other. So far those cannot be explained by the simplified diffusion
models. As stated by \citet{kochukhov-2004}, a more complex interaction
of rotation, magnetic field, and mass loss might be the solution for that.

Reconstruction of the true magnetic geometry of \thestar\ by means of magnetic Doppler imaging
(against a simplified pure dipolar geometry assumed in this work) is the next necessary step.
Such an analysis will certainly provide us with deeper insight into the connection between 
location of the magnetic poles, their strengths, and distributions of chemical elements.

\section{Conclusions}
\label{sec:conclusions}

The spectrum analysis of \thestar\ was carried out in order to determine atmospheric
parameters for Doppler imaging. By assuming a centered dipolar magnetic field,
the inclination of rotational axis to the magnetic field axis was determined to
be close to $88^\circ$. Longitudinal magnetic field variation implies a dipolar
field with a polar strength of $B_{\rm p}\approx7$~kG.

Horizontal inhomogeneities were detected
and investigated using the Doppler imaging code INVERS12.
However, this was done without direct incorporation of the magnetic field into
the spectrum synthesis calculation. The resulting maps were summarized in
three groups corresponding to the association of the surface patterns and their
relation to the magnetic field.

Lithium, O, Mn, Pd and Nd are mainly
concentrated in the area of the magnetic poles and depleted in the regions
around the magnetic equator. The high abundance spots of Si, La,
Ce, Eu, and Gd are located between the magnetic
poles and the magnetic equator. Except for La, which is clearly
depleted in the area of the magnetic poles, no obvious other correlation with
the magnetic field has been found for these elements. Ca, Cr,
and Fe show enhancements at the magnetic poles, but also along the rotational
equator. The intersection between the magnetic and the rotational
equator constitutes an exception, especially for Ca and Cr, which
are depleted in that region.

No obvious correlation between theoretical predictions of diffusion in CP
stars and the abundance patterns could be found. This is likely attributed to a lack of up-to-date theoretical
models. For elements like O or Ca,
belts of enhanced abundance around the magnetic equator are predicted,
but not present in all CP stars. Unlike some CP stars, including \thestar\ analyzed in this paper,
O is concentrated around the magnetic poles.

Finally, to complete the investigation of the abundance structures in the atmospheric layers of the \thestar,
further investigations will be pointed toward the accurate reconstruction of the surface magnetic field configuration 
and expected vertical abundance gradients will be investigated and modeled.

\begin{acknowledgements}
We thank Markus Hareter for his help with error estimates of the rotation period.
This work was supported by the following grants: Deutsche Forschungsgemeinschaft (DFG)
Research Grant RE1664/7-1 to DS. TL recieved financial contributions from the Austrian Agency for International Cooperation 
in Education and Research (WTZ CZ-10/2010) and WW from the Austrian Science Funds (P 22691-N16).
NAD thanks support of the Saint-Petersburg University, Russia,
under the Project 6.38.73.2011.
TR acknowledges RFBR grant 09-02-00002, and the Russian Federal Agency on Science and Innovation
grant Nr. 02.740.11.0247 for partial financial support.
OK is a Royal Swedish Academy of Sciences Research Fellow supported by grants
from the Knut and Alice Wallenberg Foundation and the Swedish Research Council.
We also acknowledge the use of electronic databases (VALD, SIMBAD, NASA's ADS).
\end{acknowledgements}


\end{document}